\documentclass[prl,twocolumn,showpacs]{revtex4}
\usepackage[french]{babel}
\usepackage{amsmath}
\usepackage{amssymb}
\usepackage[T1]{fontenc}
\usepackage{graphicx}
\usepackage{epsfig}

\begin{document}

\title{Measurement of the radiative and non-radiative decay rates of single CdSe nanocrystals through controlled modification of their spontaneous emission}

\author{X. Brokmann$^{1}$, L. Coolen$^{1}$}
\author{M. Dahan$^{1}$}
\email{maxime.dahan@lkb.ens.fr}
\author{J.P. Hermier$^{1,2}$}
\email{hermier@lkb.ens.fr}

\affiliation{$^{1}$Laboratoire Kastler Brossel, Ecole normale supérieure,
CNRS et Université Pierre et Marie Curie, 24 rue Lhomond, 75231 Paris Cedex 05, France. \\
$^{2}$Laboratoire Matériaux et Phénomènes Quantiques,
Université Denis Diderot, 2 place Jussieu, 75251 Paris
Cedex 05, France.}

\begin{abstract}
We present a simple method to measure the radiative and non-radiative recombination rates of individual fluorescent emitters at room temperature. By placing a single molecule successively close and far from a dielectric interface and simultaneously measuring its photoluminescence decay and its orientation, both the radiative and non-radiative recombination rates can be determined. For CdSe nanocrystals, our results demonstrate that the fluorescence quantum efficiency, determined at the single molecule level, is 98\% in average, far above the value expected from conventional ensemble experiments. The bi-dimensionnal nature of the transition dipole is also directly evidenced from a single particle measurement.
\end{abstract}


\maketitle

Among the variety of nanoscopic fluorescent emitters, colloidal CdSe nanocrystals have attracted growing attention. The spectral properties and the photostability at room temperature of these quantum dots (QDs) make them promising light sources for a wide range of applications, including quantum cryptography \cite{Mich00}, optoelectronic devices \cite{Coe02} or biological detection \cite{Alivisatos04}. The study of their optical properties has greatly benefited from the advent of single-molecule techniques. When observed individually, QDs have indeed proved to be more complex than ensemble-averaged studies could have infered. Striking phenomena such as fluorescence intermittency \cite{Nirm96}, spontaneous spectral shifts \cite{Emp96} and fluorescence-lifetime fluctuations \cite{Schl02} have been observed. The causes of these processes are not well established yet and, in light of the number of potential applications of QDs, a more detailed understanding of their emission properties remains necessary.

In this Letter, we present a simple method to obtain new information on the time-resolved fluorescence properties of single nanocrystals. Inspired by experiments on the fluorescence of layers of europium atoms placed on a dielectric interface \cite{Kun80}, we modified the electromagnetic interaction between a single CdSe QD and its optical environment while measuring the photoluminescence (PL) decay rate of the particule. It resulted in a controlled modification of the spontaneous emission from which both the radiative and non-radiative decay rates $(k_{\mathrm{rad}}$ and $k_{\mathrm{nrad}}$) were determined at the single QD level. From these data, we derived for the first time a single-particle measurement of the emitting state \emph{quantum efficiency} $Q=k_{\mathrm{rad}}/(k_{\mathrm{rad}}+k_{\mathrm{nrad}})$, i.e. the probability for a QD in its lowest-excited state to relax by emitting a photon.

\begin{figure}
\includegraphics[width=8.5cm]{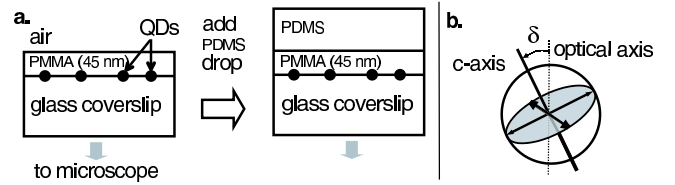}
\caption{Experimental layout. a) The PL decay rate is first measured when QDs are
close ($d \approx$ 45 nm) to a glass/air interface (left panel). A PDMS drop is then added on the sample to deplace the glass/air interface far from the emitter (right panel). b) Degenerate structure of the dipole of a single QD.}
\label{fig1}
\end{figure}

In our experiments, CdSe/ZnS core-shell nanocrystals (1.7 nm core radius, $\lambda_{0} = 560$ nm peak emission) were spin-coated on a glass coverslip and covered by a layer of polymethyl(methacrylate) (PMMA) (Fig.\:\ref{fig1}). Profilometry measurements indicated that the PMMA layers were uniform over the sample with a thickness $d=45\pm3$ nm. Since the glass coverslip and the PMMA layer have a similar index of refraction ($n_{\mathrm{glass}}$=1.52 and $n_{\mathrm{PMMA}}$=1.49), the QDs were considered as embedded in a homogenous medium with refractive index $n=1.50$ and located at a distance $z = d$ from a glass/air interface. To modify the QD fluorescence properties, a thick (>200 $\mu$m) drop of polydimethylsiloxane (PDMS, $n_{\mathrm{PDMS}}$=1.45) could be placed at any time on the sample. The effect of this additional optical medium is to remove the glass/air interface and to place the QDs in an unbounded medium with an index $n$ approximately equal to 1.5 \cite{noteH}. The experiment thus consists in measuring the PL decay rate $k = k_{\mathrm{rad}}+ k_{\mathrm{nrad}}$ before and after deposition of the PDMS droplet. Since the QD is protected by the layer of PMMA, its immediate environment is not affected by the PDMS and $k_{\mathrm{nrad}}$ can be assumed to remain constant. Consequently, changes in $k$ result only from modifications of $k_{\mathrm{rad}}$.

The key point is that $k_{\mathrm{rad}}$ depends on the optical environment of the emitters. For an emitter in an unbounded dielectric medium with refractive index $n$, the Fermi golden rule indicates that the radiative decay $k_{\mathrm{rad}}$ is enhanced by a factor $n$ compared to $k_{\mathrm{rad}}^{vac}$, its value in vacuum. This result holds as long as the emitter is at a distance $z>\lambda$ from any dielectric interface. When $z<\lambda$, detailed theoretical investigations \cite{Luk77} indicate that the spontaneous emission rate $k_{\mathrm{rad}}$ strongly depends on $d$, on the dipole orientation and on the refractive indexes at the interface. Experiments performed on layers of europium atoms \cite{Kun80} successfully resolved the predicted $z$ dependence of $k_{\mathrm{rad}}$. Recently, single-molecule experiments in far field microscopy also observed the expected orientational dependence of $k_{\mathrm{rad}}$ at a given distance \cite{Mac96,Krei02}. In the following, we note $\alpha = k_{\mathrm{rad}}(d)/k_{\mathrm{rad}}^{\infty}$ the ratio of the radiative recombination rate of the emitter at distance $z=d$ from the interface to $k_{\mathrm{rad}}^{\infty}=n k_{\mathrm{rad} }^{vac}$, its value in an unbounded medium.

\begin{figure}[t]
\includegraphics[width=7 cm]{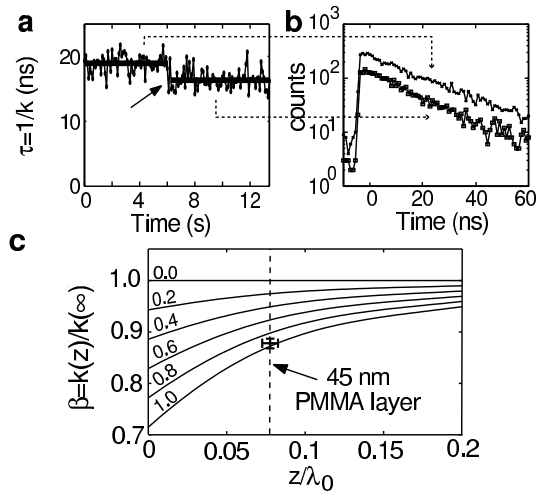}
\caption{Ensemble measurement of the quantum efficiency. a) Time trace of PL lifetime $1/k$ (time bin : 100 ms). The arrow indicates the time at which the PDMS droplet was added. b) Monoexponential PL decay before and after the deposition of PDMS (time bin: 100 ms). c) Solid lines : $\beta = k(z)/k^{\infty}$  calculated for uniformly oriented dipoles as a function of the distance $z$ from the interface and for values of $Q$ equal to 0, 0.2, 0.4, 0.6, 0.8 and 1. The experimental data yield a value of $Q$ greater than 0.9.}
\label{fig2}
\end{figure}

In general, approaching an emitter close to an interface with a medium having a lower refractive index results in a decrease of $k_{\mathrm{rad}}$, i.e $\alpha <1$ \cite{Luk77,alpha}. In our case, adding a thick droplet of PDMS on the sample amounts to changing the emitter/interface distance from $z=d$ to $z=\infty$ and thus increasing $k_{\mathrm{rad}}$ by a factor $\alpha^{-1}$. This factor $\alpha$ translates into a change $\beta = k(d)/k^{\infty}$ of the PL decay rate (Fig.\:\ref{fig2}) which is experimentally obtained by successively measuring $k(d) = \alpha k_{\mathrm{rad}}^{\infty} + k_{\mathrm{nrad}}$ and $k^{\infty} = k_{\mathrm{rad}}^{\infty} + k_{\mathrm{nrad}}$. These measurements, combined to the computation of $\alpha$, directly provide the values of
$k_{\mathrm{rad}}^{\infty}$, $k_{\mathrm{nrad}}$ and the quantum efficiency of the emitter far from the interface $Q=k_{\mathrm{rad}}^{\infty}/(k_{\mathrm{rad}}^{\infty} + k_{\mathrm{nrad}})=(\beta-1)/(\alpha-1)$.

The method was first tested on an ensemble of QDs excited at 400 nm by a pulsed laser diode (LDH400 PicoQuant). The fluorescence photons were collected by a  microscope objective (1.4 NA, 100$\times$ Olympus Apochromat) and sent to an avalanche photodiode followed by a time-resolved photon counting card (TimeHarp200, PicoQuant). The PL intensity and decay were recorded before and after the PDMS droplet deposition in the time-tagged time-resolved (T3R) acquisition mode. The PL decay curve in each 100 ms time bin was found to be monoexponential \cite{Baw04} and was fitted accordingly using a Gauss-Newton method (Fig.\:\ref{fig2}a-b). The droplet deposition caused an increase of the PL decay rate $k$ by a factor $\beta^{-1}= 1.14$ ($k^{\infty}= 55 \mu$s$^{-1}$).

The computation of $\alpha$, necessary to deduce the value of $Q$, requires in principle to take into account the nature of the transition dipole of QDs. For CdSe nanoparticles, the two degenerate emitting excitonic states have an angular momentum projection $\pm \hbar$ along the c-axis of their  hexagonal crystal structure \cite{Efr96}. In contrast to most fluorophores which have linear emission dipoles, the QD transition dipole is therefore circular (2D degenerate) and located in a plane perpendicular to its c-axis, as demonstrated both at low and room temperature \cite{Emp99}.

Using expressions given in \cite{Luk77}, we numerically computed $\alpha$ for a dipole perpendicular or parallel to the interface and obtained $\alpha_{\perp} = 0.61$ and $\alpha_{\parallel}$=1.00 for an emitter at a distance $d= 45$ nm from the interface \cite{Discussion}. However, averaging $\alpha$ over uniformly oriented 1D or 2D dipoles leads to the same value $<\alpha> =(\alpha_{\perp}+ 2\alpha_{\parallel})/3 = 0.87$. An ensemble measurement can not discriminate between 1D or 2D dipoles but it can nevertheless provide an estimate for $Q$. For this sample, we deduced  $<k_{\mathrm{rad}}^{\infty}>= (\beta - 1)k^{\infty}/(<\alpha> - 1)= 52 \mu$s$^{-1}$, $<k_{\mathrm{nrad}}>=3 \mu$s$^{-1}$ and $Q=$0.94 $\pm0.15$, the uncertainty resulting from the uncertainties over both $d$ and $\beta$ ($\beta=0.88\pm0.01$).

\begin{figure}
\includegraphics[width=7cm]{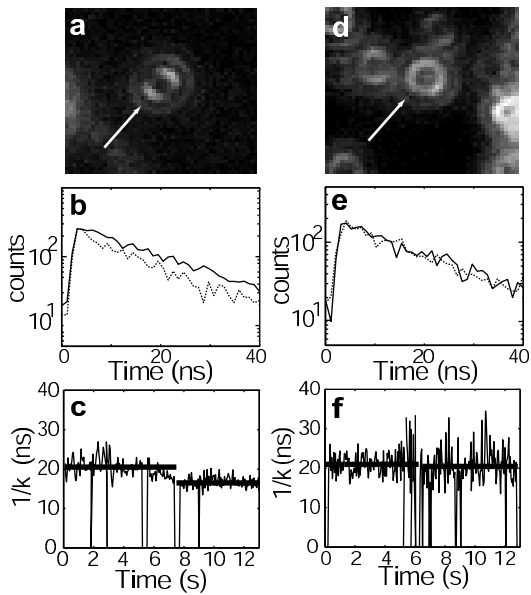}
\caption{Single-QD measurements of the quantum efficiency. a) defocused image for an horizontal QD. b) PL decay before (solid line) and after (dotted line) the droplet deposition (time bin: 1 s). c) Trace of the PL decay time (time bin: 50 ms). The horizontal solid line indicates the average lifetime measured before and after adding the droplet deposition showing a reduction of $1/k$ by a factor $\beta = 0.81$. d) defocused image for a vertical QD. e-f) PL decay time and trace of the lifetime. $1/k$ was unchanged when the PDMS was added ($\beta=0.99$).}
\label{fig3}
\end{figure}

To avoid ensemble averaging and to probe the orientational dependence of $\alpha$ and $\beta$, further investigations were conducted on single emitters. For each QD, its orientation was determined by defocused imaging, a technique previously introduced for linear dipoles \cite{Jas97}. This technique consists in moving the microscope objective, originally in focus, toward the sample by 1 $\mu$m and then recording a fluorescence image. This defocused image is an expression of the emitter radiation pattern, which for
CdSe QDs results from the sum of the patterns corresponding to two orthogonal 1D dipoles with equal amplitudes. From the anisotropy of the defocused image \cite{Jas97}, one unambiguously deduces the 3D orientation of the nanoparticle and, in particular, the value of the angle $\delta$ between the QD c-axis and the sample plane (with an accuracy of $\sim 10^{\circ}$) (Fig.\:\ref{fig1}c) \cite{Brok04}. For a
an `horizontal' QD with its c-axis parallel to the coverslip ($\delta = 90^{\circ}$), the defocused spot has an axial symetry along a direction perpendicular to the c-axis (Fig.\:\ref{fig3}a). In contrast, a `vertical' QD ($\delta = 0^{\circ}$) leads to a defocused pattern with a circular symetry around the optical axis (Fig.\:\ref{fig3}d).

In single-QD measurements, the fluorescence intensity was recorded for each particle before and after the deposition of PDMS. To account for the fluorescence intermittency of individual QDs, the PL decay rate was calculated during bright periods for which the intensity is high. Here again the decay was well described by a monoexponential curve, consistent with previous observations  (Fig.\:\ref{fig3}b-e) \cite{Baw04}. We first studied the change of PL decay rate for horizontal QDs for which $\alpha= (\alpha_{\perp}+\alpha_{\parallel})/2=0.8$. Upon droplet deposition, the mean $k$ increased by a factor $\beta^{-1}=1.24$ (Fig.\:\ref{fig3}c). For this QD, we therefore deduce $Q= 0.95\pm0.15$. When averaged over 8 horizontal QDs, the mean value of $Q$ was 0.97 with a standard deviation (s.d.) 0.08.

For vertical QDs, the deposition of a PDMS droplet did not modify the PL decay rate ($\beta=0.99$ on Fig.\:\ref{fig3}e-f). This observation is consistent with both linear components of the 2D dipole being aligned within the sample plane. In this case, $\alpha=\alpha_{\parallel}$ is very close to 1 and no significant change in $k$ is expected. Although measurements for vertical QDs do not allow for a measurement of $Q$, they provide a simple demonstration of the 2D nature of the emitting dipole. For a linear dipole, a circularly symetric defocused pattern is obtained when the dipole is perpendicular to the sample plane. In this case, $\alpha=\alpha_{\perp}$ is maximum and $\beta = 1 + Q(\alpha-1)$ should be significantly lower than 1, in contrast with the experimental result. Previous experiments demonstrating the 2D nature of the transition dipole in CdSe nanocrystals were based on polarization measurements \cite{Emp99}. The distribution of the polarization anisotropy in the QD emission was compared to the one for DiI molecules which have a linear dipole. Assuming that the orientations of the particles were uniformly distributed, disagreement between the two distributions led to the conclusion that QDs had a 2D degenerate emission dipole. In comparison, our method does not require any assumption on the orientational distribution and the nature of the transition dipole is directly deduced from a measurement on a single QD.

\begin{figure}
\includegraphics[width=7.5cm]{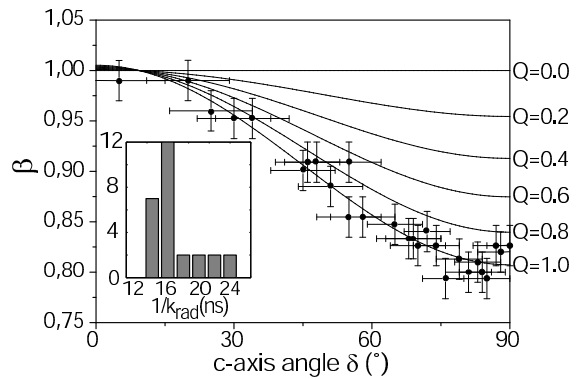}
\caption{Orientational dependance of $\beta$ observed on 27 single QDs ($\circ$). The solid lines correspond to $\beta$ calculated for different values of $Q$ at $d$=45 nm. Inset : histogram of the radiative lifetime.}
\label{fig4}
\end{figure}

Further data were taken to probe the value of $Q$ for QDs with an orientation intermediate between $\delta = 0^{\circ}$ and $\delta=90^{\circ}$. The results indicated that all individual QDs had a very high quantum efficiency (Fig.\:\ref{fig4}). The homogeneity of the measured quantum efficiencies was actually consistent with the value of $Q$ inferred from the ensemble measurement. When averaged over 21 particles (all with $\delta>45^{\circ}$ to allow a precise determination of $Q$), the mean quantum efficiency was equal to 0.98 (s.d. 0.11).

The dependence of $Q$ with the core radius was also investigated using other samples of CdSe/ZnS QDs having a peak emission at 535, 585, 600 and 610 nm (corresponding to core radii of 1.4, 1.9, 2.2 and 2.4 nm respectively). For each of these samples, we performed an ensemble measurement using the procedure described above and values of $Q$ higher than 95\% were systematically found. This observation confirms that quantum efficiencies close to 1 are a common property of CdSe/ZnS QDs, as recently suggested (but not demonstrated) on the basis of the homogeneity of the single-QD lifetimes \cite{Baw04}.

The brightness of a QD sample is conventionally characterized with a cuvette measurement of the \emph{quantum yield} $\chi$. In these measurements, $\chi$ is obtained by comparing the PL intensity between a solution of QDs and a solution (having the same absorption) of reference molecules (such as rhodamine in ethanol). For our samples, $\chi$ was on the order of 30 \%, comparable to values typically reported for CdSe/ZnS core-shell quantum dots \cite{Hines96} but well below the values of $Q$ that we measured. Several factors can account for this discrepancy. Firstly, $\chi$ corresponds to the probability of emitting a fluorescence photon given that a pump photon have been absorbed. It can thus differ from $Q$ if the electron-hole pair can recombine non-radiatively during its rapid transfer from the highly-excited state after pumping to the lowest-excited (emitting) state. Secondly, in a cuvette measurement the fluorescence intermittency of individual QDs is not taken into account and the fraction $F$ of QDs that indeed emit photons is not evaluated. All the QDs are assumed to be fluorescent and, as a result, the value of $\chi$ tends to be underestimated. The fraction $F$ has been determined using different experimental techniques \cite{Ebenstein02,Brokmann03,Webb04} and values of $F$ on the order of 20-50 \% have been reported, possibly explaining the difference between $Q$ and $\chi$.

Our findings unambiguously demonstrate that the recombination process of the emitting state in CdSe QDs is almost entirely radiative, reinforcing the image of QDs as artificial atoms. From our data, one deduces $k_{\mathrm{nrad}} \approx 1\;\mu$s$^{-1}$ which, considering the accuracy of our single-molecule measurements, should only be considered as an upper bound. This value was compared to the result of a simple particle-in-a-box model in which the first quantum confined state is coupled by resonant-tunnelling to a surface trap state \cite{Dabbousi97}. For a 1.7 nm core radius CdSe nanocrystal coated with a 1 nm ZnS shell, $k_{\mathrm{nrad}}$ as low as $1\;\mu$s$^{-1}$ implies that no resonant trap can be located closer than the outer ZnS surface of the QD. This suggests that the ZnS passivation layer efficiently removes all non-radiative traps from the CdSe core surface.

Finally, our experiments provide a value of 17.0 (s.d. 2.0) ns for the average radiative lifetime $1/k_{\mathrm{rad}}^{\infty}$ (Fig.\:\ref{fig4}), corresponding to a lifetime of 25.5 ns in vacuum. We calculated the radiative lifetime at room temperature based on the electronic structure of CdSe nanoparticles described in \cite{Efr96} and found, for spherical QDs emitting at 560 nm in vacuum, a value of 31 ns, in reasonable agreement with the experimental result. This discrepancy might be explained by the strong theoretical dependence of $1/k_{\mathrm{rad}}^{\infty}$ on both the ellipticity and the emission wavelength of the particle.

In conclusion, we used a simple method to modify in a controlled manner the spontaneous emission rate of individual QDs. Combined with a determination of the QD orientation, this method yielded the values of both the radiative and non-radiative decay rates of individual nanocrystals. The resulting quantum efficiency was found to be close to unity for all the QDs. Our method is not restricted to QDs and should find applications for the study of other fluorescent species.

We are grateful to H. Rigneault for valuable discussion. This work has been supported by the "S4P" project from the European Union IST/FET/QIPC program and by ACI Cryptologie from Ministere de la Recherche.


\begin{thebibliography}{99}
\bibitem{noteH}{Numerical simulations confirmed that the slight differences between the values of $n_{\mathrm{glass}}$, $n_{\mathrm{PMMA}}$ and $n_{\mathrm{PDMS}}$ could be neglected.}
\bibitem{alpha}{For some particular values of $d$ and of the dipole orientation, $\alpha$ can actually be slightly greater than 1 (see Fig.\:\ref{fig4} and \cite{Luk77}).}
\bibitem{Discussion}{The change in lifetime can be viewed as the effect on the dipole of its own field reflected by the dielectric interface. While this effect is very modest for a parallel dipole ($\alpha_{\parallel} = 1.004$), it is more significant for a perpendicular dipole and, the field being out-of phase with the dipole for $d= 45$ nm, the radiative lifetime is increased close to the interface ($\alpha_{\perp}=0.613 < 1$) \cite{Luk77}.}
\end{thebibliography}
\end{document}